\documentclass[aip,jap,reprint,floatfix]{revtex4-2}
\usepackage{graphicx}
\usepackage{amsmath,amssymb,amsfonts}
\usepackage{amsthm}

\begin{document}

\title{Benchtop Momentum-Resolved Phonon Spectroscopy: Unlocking Lattice Dynamics via X-ray Diffuse Scattering}

\author{Paulo H. V. da Silva}
\author{Tarsis M. Germano}
\author{Rafaela F. S. Penacchio}
\author{S\'ergio L. Morelh\~ao}
\affiliation{Institute of Physics, University of S\~ao Paulo, S\~ao Paulo, SP, Brazil}

\date{\today}

\begin{abstract}
Momentum-resolved phonon spectroscopy is essential for understanding the thermal, electronic, and structural properties of materials, yet it traditionally necessitates access to large-scale synchrotron or neutron facilities. We report a paradigm shift by demonstrating the accurate measurement and quantitative analysis of Thermal Diffuse Scattering (TDS) patterns using a commercially available benchtop Small/Wide-Angle X-ray Scattering (SAXS/WAXS) instrument. By utilizing a low-energy ($\sim$8 keV) X-ray source and a custom-integrated vacuum geometry, we acquired characteristic TDS patterns in silicon (111) and (100) slabs. We demonstrate that these benchtop images contain sufficient information to extract interatomic force constants (IFCs) through global optimization algorithms, achieving a fitting accuracy comparable to those obtained from simulated data with statistical noise. Crucially, this capability provides access to phonon information across the entire Brillouin Zone—a critical component of lattice dynamics unavailable via standard laboratory techniques like first-order Raman spectroscopy. This low-barrier approach makes synchrotron-level lattice dynamics characterization accessible to virtually any research laboratory, enabling new avenues for the study of intrinsic strain and anisotropic lattice dynamics in emerging material systems.
\end{abstract}

\maketitle

\section{Introduction}\label{sec1}

It is well know that phonons, the elementary quanta of lattice vibrations, strongly influence the thermodynamic and transport properties of solids. Over the past decade, a substantial number of  studies have further reinforced the importance of understanding lattice dynamics in modern materials science. For instance, \citeauthor{Huang2024} (\citeyear{Huang2024}) demonstrated that power conversion efficiency can be dramatically enhanced in two-dimensional van der Waals solar cells through modulation of lattice vibrations \cite{Huang2024}, while suppression of electron-phonon coupling has emerged as a promising strategy for reducing non-radiative energy loss in organic solar cells \cite{Luo2026}. Beyond energy-related applications, the fast-developing field of topological phononics holds promise for applications in quantum information science \cite{Liu2020,Xu2024}. In non-centrosymmetric materials, topological and chiral phonons have gained prominence as vital degrees of freedom for understanding physical processes \cite{Zhang2025}. Lattice vibrational signatures are also essential for characterizing phenomena such as polaron formation in anharmonic semiconductors, where charge carrier screening and extended lifetimes are critically mediated by dynamically disordered phonons \cite{Wang2022}. However, to date, the investigation of phonon dispersion relations remains largely theoretical, often relying on high-throughput density functional theory (DFT) calculations without direct experimental validation of the underlying interatomic force constants (IFCs) \cite{Aouissi2006}. The latter is particularly problematic because higher-order IFCs, required to describe anharmonicity and thermal transport, are highly sensitive to the choice of exchange–correlation functional used in the calculations. Such sensitivity leads to significant discrepancies among theoretical predictions, creating a demand for direct experimental determination of bulk IFCs \cite{Zhou2023}.

While laboratory-based Raman spectroscopy is the most accessible tool for probing phonons, it is fundamentally restricted by selection rules to the Brillouin zone center ($\Gamma$-point), leaving the vast majority of momentum-dependent lattice dynamics experimentally unverified \cite{Adu2012}. Among momentum-resolved spectroscopies, Inelastic Neutron Scattering (INS) \cite{Brockhouse1959,Lovesey1977,Kim2020} and Inelastic X-ray Scattering (IXS) \cite{Burkel2000,Miao2018,Baron2020} have long been the ``gold standards" for mapping energy-resolved phonon dispersions. However, both techniques are logistically tethered to large-scale facilities, such as nuclear reactors or synchrotrons. Beyond these logistical hurdles, INS is severely limited by the requirement for large sample volumes, often necessitating single crystals on the order of cubic centimeters. Resonant Inelastic X-ray Scattering (RIXS) \cite{Ament2011,Groot2024,Thomas2025,Zinouyeva2026} is a powerful high-sensitive probe of elementary excitations, but its reliance on high-brightness tunable X-ray sources limits its widespread use. A similar limitation applies to state-of-the-art techniques based on X-ray Thermal Diffuse Scattering (TDS), whether in transmission or reflection geometries\cite{Holt1999,Xu2005,Mei2015}, which remain confined to synchrotron sources. Finally, electron-based probes, like Electron Energy Loss Spectroscopy (EELS) \cite{Elgvin2026,Li2023} and electron diffuse scattering \cite{Seiler2021}, provide exceptional momentum resolution for surface and thin-film phonons, but lacking the penetration depth required for bulk material characterization. 

In this work, we report accurate measurements and quantitative analysis of TDS patterns collected using a commercially available benchtop X-ray instrument. Though synchrotron-based transmission and reflection TDS have established the feasibility of full-zone mapping, we bring this capability to the laboratory setting using a low-energy ($\sim$8 keV) source and integrated vacuum geometry. We show that these benchtop diffraction images contain sufficient information to extract IFCs through global optimization algorithms, achieving a fitting accuracy comparable to synchrotron-level data. We demonstrate this approach using Si as model material system. We anticipate that the methodology presented here has the potential to democratize the access to lattice-dynamics characterization, enabling routine studies of intrinsic strain and anisotropic vibrations in new materials.

\section{Results}\label{sec2}

\begin{figure*}
    \centering
    \includegraphics[width=\linewidth]{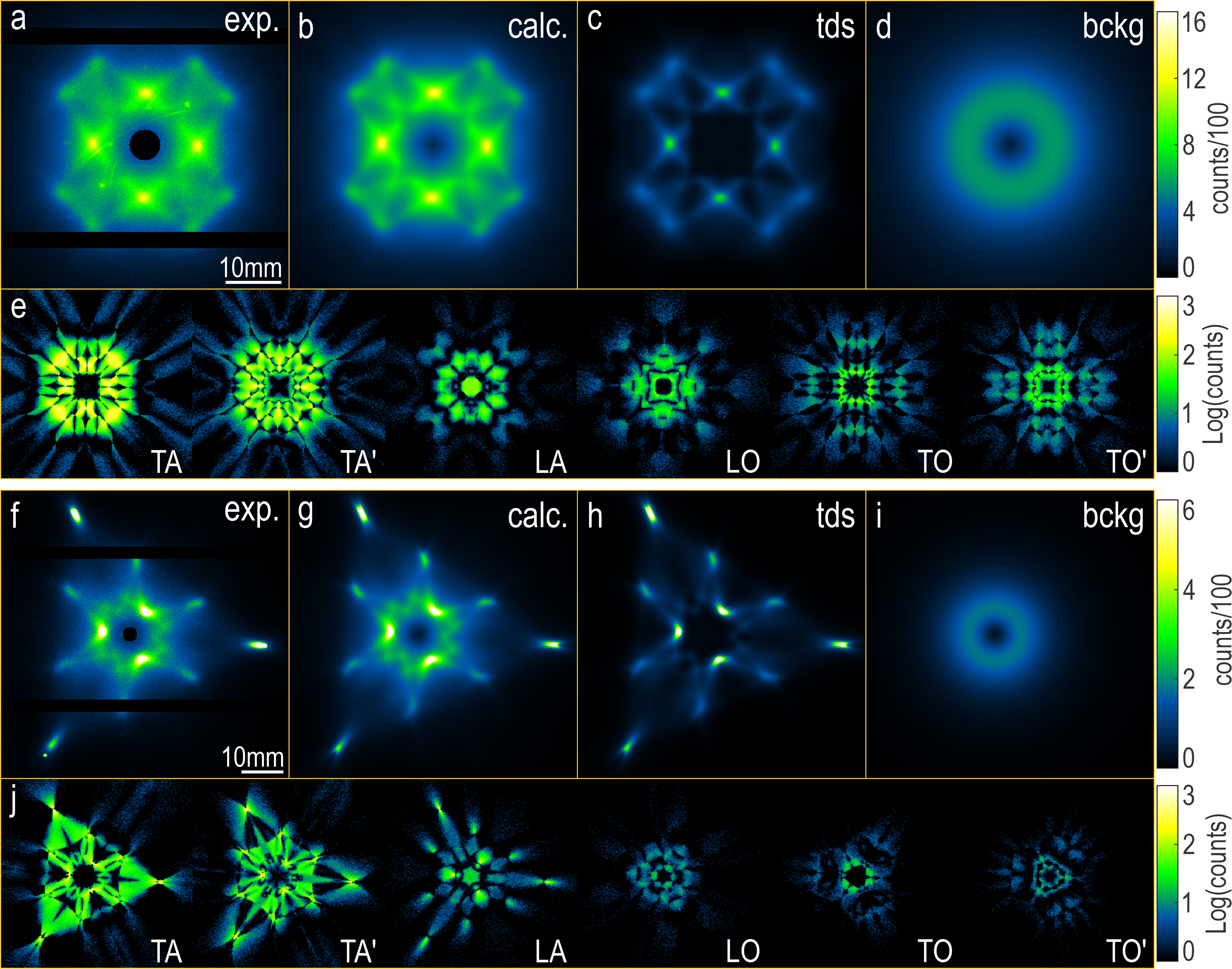}
    \caption{Comparison between experimental and simulated thermal diffuse scattering (TDS) patterns for silicon single crystals. Top panel (a–e) corresponds to the Si(100) sample, and bottom panel (f–j) corresponds to the Si(111) sample. Panels display: (a, f) experimental diffuse scattering intensity; (b, g) total simulated pattern obtained from the sum of the calculated TDS and the isotropic background; (c, h) calculated TDS pattern derived from the refined Born–von Kármán IFCs; (d, i) isotropic background profile modeled via an asymmetric split pseudo-Voigt function representing disordered/amorphous scattering contributions; and (e, j) mode-resolved decomposition displaying the individual calculated TDS contributions for each of the six acoustic (TA, TA$^\prime$, LA) and optical (LO, TO, TO$^\prime$) phonon branches. Measurements/calculations were performed using Cu-$K_\alpha$ radiation with a sample-to-detector distance of approximately 1.3\,cm.}
    \label{fig:tdsnew}
\end{figure*}

Experimental and calculated TDS patterns from silicon single-crystal samples in two distinct crystallographic orientations are presented in Fig.~\ref{fig:tdsnew}. In both cases, a direct comparison between the experimental data (Figs.~\ref{fig:tdsnew}a and \ref{fig:tdsnew}f) and simulations (Figs.~\ref{fig:tdsnew}b and \ref{fig:tdsnew}g) demonstrates remarkable agreement, clearly resolving all essential phonon-related features, such as the characteristic ``butterfly'' intensity distributions \cite{Holt1999,Xu2005}. The calculated patterns comprise two components: (\textit{i}) the theoretical TDS patterns (Figs.~\ref{fig:tdsnew}c and \ref{fig:tdsnew}h) derived from the Born--von K'arm'an (BvK) force-constant model of lattice dynamics \cite{Herman1959,Wang1993}, and (\textit{ii}) an isotropic background (Figs.~\ref{fig:tdsnew}d and \ref{fig:tdsnew}i) accounting for diffuse scattering from disordered or amorphous contributions. The observed TDS features arise from multi-phonon scattering processes and exhibit high sensitivity to the dispersion anisotropy of both acoustic and optical branches. This is demonstrated by decomposing the calculated patterns into their mode-specific contributions in Figs.~\ref{fig:tdsnew}e and \ref{fig:tdsnew}j, where the transverse acoustic ($\mathrm{TA, TA^\prime}$) branches dominate the high-contrast butterfly wings, while the longitudinal acoustic ($\mathrm{LA}$) and optical ($\mathrm{TO, TO^\prime, LO}$) branches form a smoother intensity background and interconnecting diffuse bridges.

\begin{table}
    \centering
    \begin{tabular}{crrr}
\hline\hline
Symbol  & Si (100) & Si (111) & INS\\
\hline
 $   \alpha$ &  55.21834 ($\pm$1.81\%) &  54.24360 ($\pm$1.84\%) &  55.00246 \\ 
 $    \beta$ &  40.47607 ($\pm$2.47\%) &  41.54454 ($\pm$2.41\%) &  39.74917 \\ 
 $    \mu_2$ &   1.55377 ($\pm$6.44\%) &   4.23094 ($\pm$2.36\%) &   2.97828 \\ 
 $    \nu_2$ &   0.84381 ($\pm$1.19\%) &   5.04983 ($\pm$1.98\%) &   3.11542 \\ 
 $ \delta_2$ &   0.23909 ($\pm$4.18\%) &   2.01899 ($\pm$4.95\%) &   1.61676 \\ 
 $\lambda_2$ &  -4.35901 ($\pm$2.29\%) &  -6.37355 ($\pm$1.57\%) &  -6.75533 \\ 
 $    \mu_3$ &  -1.42321 ($\pm$7.03\%) &  -0.42790 ($\pm$2.34\%) &  -0.56993 \\ 
 $    \nu_3$ &   0.08726 ($\pm$1.15\%) &   0.78701 ($\pm$1.27\%) &   1.06313 \\ 
 $ \delta_3$ &  -0.46567 ($\pm$2.15\%) &  -0.56031 ($\pm$1.78\%) &  -0.50542 \\ 
 $\lambda_3$ &   0.04869 ($\pm$2.05\%) &  -0.08469 ($\pm$1.18\%) &  -0.08596 \\ 
 $    \mu_4$ &   1.07531 ($\pm$9.30\%) &   0.70801 ($\pm$1.41\%) &   0.42535 \\ 
 $\lambda_4$ &  -0.15658 ($\pm$6.39\%) &  -0.04077 ($\pm$2.45\%) &  -0.22121 \\ 
 $    \mu_5$ &   0.12308 ($\pm$8.12\%) &   0.13068 ($\pm$7.65\%) &   0.56957 \\ 
 $    \nu_5$ &  -0.52201 ($\pm$1.92\%) &  -0.07518 ($\pm$1.33\%) &  -0.00969 \\ 
 $ \delta_5$ &   0.17093 ($\pm$5.85\%) &   0.11128 ($\pm$8.99\%) &   0.22337 \\ 
 $\lambda_5$ &   1.82949 ($\pm$5.47\%) &   1.81293 ($\pm$5.52\%) &   2.75008 \\ 
\hline\hline
 \end{tabular}
    \caption{Interatomic force constants (IFCs), in N/m, extracted from the experimental TDS patterns (Figs.~\ref{fig:tdsnew}a and \ref{fig:tdsnew}f) for the Si (100) and Si (111) samples. Initial IFC values obtained from literature fits to INS dispersion curves \cite{Wang1993,Kim2020} are included for comparison. Values in parentheses denote parameter uncertainties expressed as a percentage, corresponding to the relative half-width of the final search interval relative to each refined IFC.}
    \label{tab:ifcvalues}
\end{table}

\begin{figure}
    \centering
    \includegraphics[width=\linewidth]{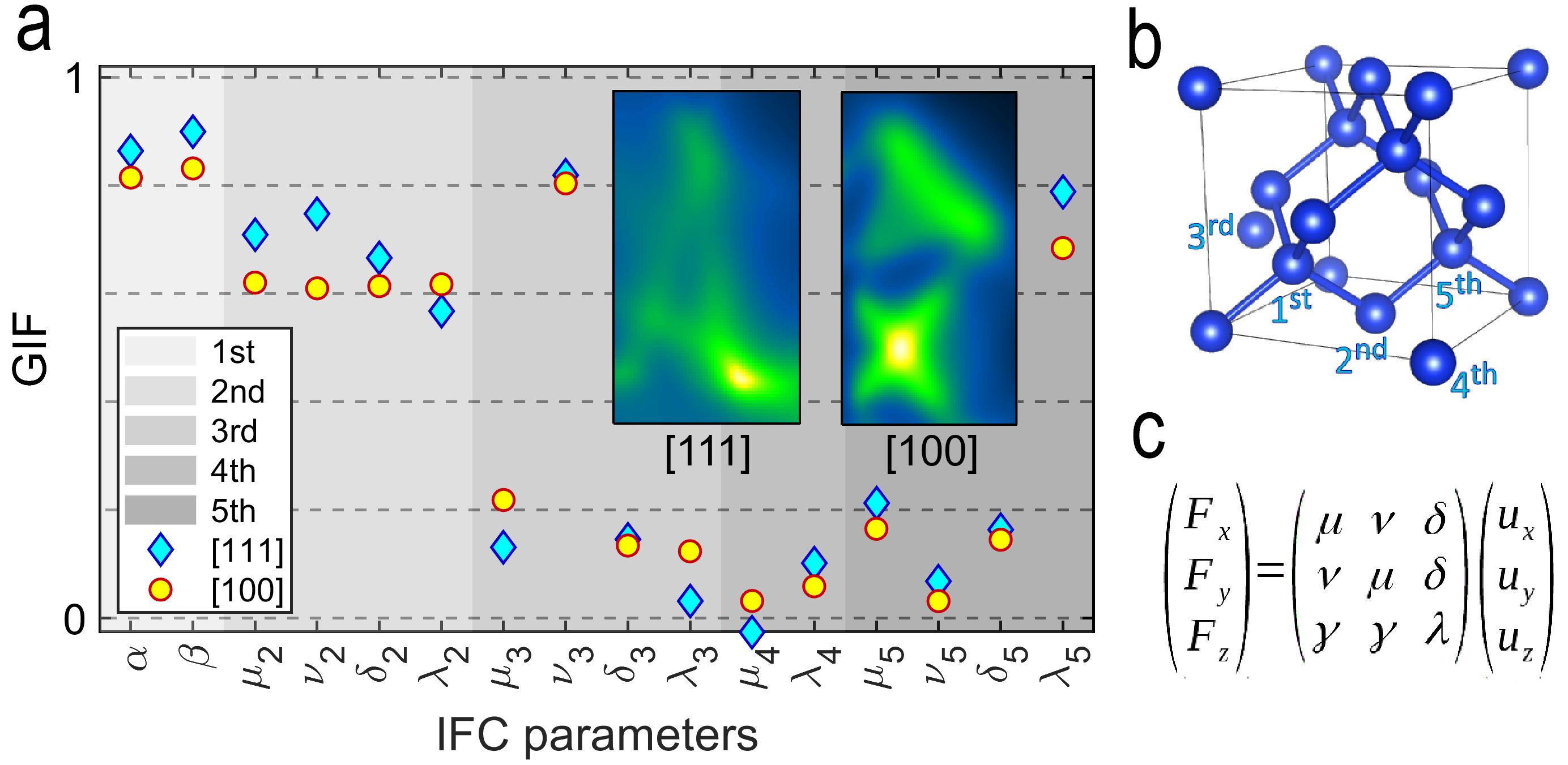}
    \caption{Parameter convergence analysis and structural framework for IFC extraction. (a) Global Impact Factor (GIF) of the IFCs calculated from the TDS patterns. The GIF quantifies the optimization algorithm's parameter convergence and sensitivity for the 1st- through 5th-nearest neighbors, with each neighbor shell indicated by areas in shades of gray and subscripts in the parameter symbols (2nd to 5th). The TDS pattern regions of interest (ROIs) for each crystal orientation are shown as insets. (b) Spatial arrangement of the nearest neighbors with respect to the Si atom at the unit cell origin. (c) Schematic matrix representation of the IFC parameter symbols, where $F_{x,y,z}$ and $u_{x,y,z}$ denote the force and atomic displacement components, respectively. For the 1st-nearest neighbor, $\mu = \lambda = \alpha$ and $\nu = \delta = \gamma = \beta$; for the 2nd, $\gamma=-\delta$; for the 3rd and 5th, $\gamma=\delta$; and for the 4th, non-diagonal elements are null ($\nu = \delta = \gamma = 0$). }
    \label{fig:GIF}
\end{figure}

IFCs up to the 5th-nearest-neighbor shell were incorporated into the harmonic Born--von K'arm'an calculations, defining the TDS intensity distributions via 16 independent parameters (listed in Table~\ref{tab:ifcvalues}). A parameter convergence analysis for IFC extraction using pattern optimization through a Differential Evolution (DE) algorithm \cite{Wormington1999} is presented in Fig.~\ref{fig:GIF}, along with the labeling convention adopted for the force-constant tensors. The isotropic background was modeled using a five-parameter asymmetric split pseudo-Voigt (PV) function. Including a scale factor for the calculated TDS pattern, a total of 22 parameters were refined. Geometric parameters defining the detector pixel array relative to the crystal frame—namely the sample-to-detector distance and sample tilts—were calibrated manually prior to optimization. Sample thicknesses were determined independently with micrometer precision ($\pm 2\,\mu\text{m}$). The optimization protocol proceeds in two stages: first, the five background parameters and the TDS scale factor are adjusted while holding the IFCs fixed at their literature values derived from neutron data (INS; Table~\ref{tab:ifcvalues}); subsequently, all IFCs are refined simultaneously alongside the scale parameter. The optimization is driven by nonlinear least-squares minimization of the log-transformed intensity residuals. The refined IFC values obtained from this procedure are summarized in Table~\ref{tab:ifcvalues} (second and third columns).

Both TDS patterns exhibit comparable sensitivity to the refined IFCs, as quantified by the Global Impact Factor ($\mathrm{GIF}$) presented in Fig.~\ref{fig:GIF}a. The impact factor is defined as $\mathrm{GIF} = 1 - \sigma / (0.288 W)$, where $\sigma$ is the standard deviation across hundreds of independent DE optimization runs and $W$ is the search-interval width (spanning one order of magnitude around the nominal IFC value; see Supplementary Information for details). Under this definition, a $\mathrm{GIF}$ of 0 reflects a uniform random distribution ($\sigma \approx 0.288 W$), whereas a $\mathrm{GIF}$ of 1 indicates ideal parameter convergence. Following the Cartesian reference frame and tensor matrix conventions illustrated in Figs.~\ref{fig:GIF}b and \ref{fig:GIF}c—where diagonal elements represent longitudinal/transverse stretching and off-diagonal elements denote orthogonal shear IFCs—our analysis reveals robust convergence ($\mathrm{GIF} \gtrsim 0.6$) for eight specific parameters: all six IFCs from the 1st- and 2nd-neighbor shells, the shear IFC $\nu_1$ in the 3rd shell, and the transverse IFC $\lambda_3$ in the 5th shell.

\section{Discussion}\label{sec3}

\begin{figure}
    \centering
\includegraphics[width=\linewidth]{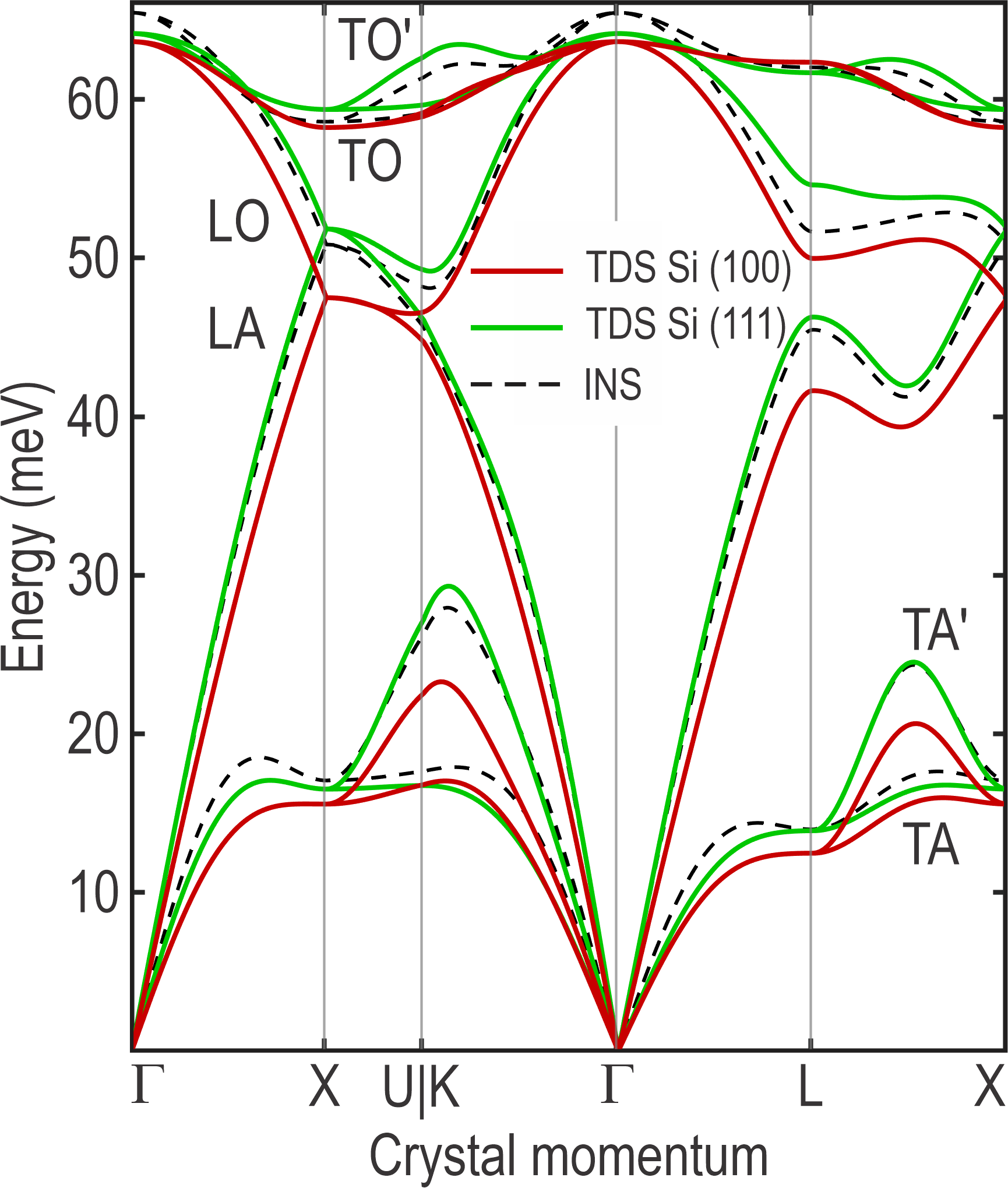}
    \caption{Phonon dispersion curves of silicon. The curves were computed by using the IFC values in Table~\ref{tab:ifcvalues}. Dashed lines are very similar to the INS data reported elsewhere \cite{Kim2020}.}
    \label{fig:dispcurves}
\end{figure}

 Phonon dispersion relations computed using the extracted IFC values are compared in Fig.~\ref{fig:dispcurves} against reference INS experimental data \cite{Wang1993,Kim2020}. Excellent agreement is achieved for the Si (111) sample across the entire Brillouin zone, demonstrating that this benchtop, low-energy vacuum setup is capable of reliably extracting higher-order IFCs. Conversely, the dispersion curves refined for the Si (100) sample exhibit slight mode softening when all 22 parameters are adjusted simultaneously during the final optimization stage. This discrepancy arises primarily from the extensive spatial overlap of the isotropic background across the Si (100) TDS pattern, unlike the Si (111) orientation where key diffuse features extend well beyond the central background scattering ring. Consequently, the closest agreement with the benchmark INS dispersions for Si (100) is achieved by holding the five PV background parameters fixed during the final IFC refinement. Furthermore, because the mode-resolved TDS intensity distributions (Figs.~\ref{fig:tdsnew}e and \ref{fig:tdsnew}j) span more than three orders of magnitude, applying a logarithmic transformation to the intensity residuals is essential for unbiased least-squares minimization across both high- and low-intensity phonon branches.

\section{Methods}\label{sec4}

\subsection{Materials}
The experimental specimens comprised thin single-crystal silicon slabs in two distinct crystallographic orientations. The (100)-oriented specimen was prepared from a commercial Si (100) wafer, polished down in-house to a thickness of $110\pm2\,\mu\text{m}$. The (111)-oriented specimen was prepared from as-grown dendritic-web silicon (DWS) ($110\pm2\,\mu\text{m}$ thick), a high-purity material developed for photovoltaic applications that naturally presents atomically flat (111) facet surfaces \cite{Morelhao2000}. Although DWS frequently develops (111) twin planes during growth, the TDS pattern in Fig.~\ref{fig:tdsnew}a was deliberately acquired from a twin-free domain, as confirmed by its pristine three-fold rotational symmetry. For comparison, patterns collected from twinned regions displaying an apparent six-fold symmetry are provided in the Appendix. Both specimens were measured in transmission geometry inside an integrated vacuum chamber to eliminate air scattering and minimize background parasitic signals.

\subsection{Geometric and Absorption Corrections}

To ensure the quantitative accuracy of the intensity fits at low sample-to-detector distances ($\sim$10–13\,mm), our numerical model incorporates corrections for both geometric and material-specific factors. We implemented pixel solid angle correction to account for the variation in the effective area subtended by each pixel as a function of the scattering angle $2\theta$, as well as polarization corrections for unpolarized X-rays. Additionally, the model accounts for photoelectric attenuation within the silicon slabs. Because the effective path length of the X-rays increases as $\sec(2\theta)$ for transmission through the slab, we applied an angle-dependent absorption correction using the linear attenuation coefficient for silicon at 8.04,keV. These refinements are essential for maintaining the fidelity of the TDS pattern further out its central region, where geometric distortion and absorption are most pronounced.

\subsection{Genetic Algorithm}
Optimization was performed using a global parallel direct-search routine based on the Differential Evolution (DE) algorithm \cite{Storn1997}, a strategy long proven effective for navigating the complex local minima characteristic of X-ray scattering data landscapes \cite{Wormington1999} and continually benchmarked as a premier tool for robust multi-parameter convergence \cite{Opara2019,Cao2024}. To enable efficient refinement of the IFCs via the genetic algorithm, the dynamical matrix calculations were accelerated using a custom C++ MEX implementation utilizing the Eigen library for high-performance linear algebra. 

\subsection{Experimental Setup}

\begin{figure}
    \centering
    \includegraphics[width=0.8\linewidth]{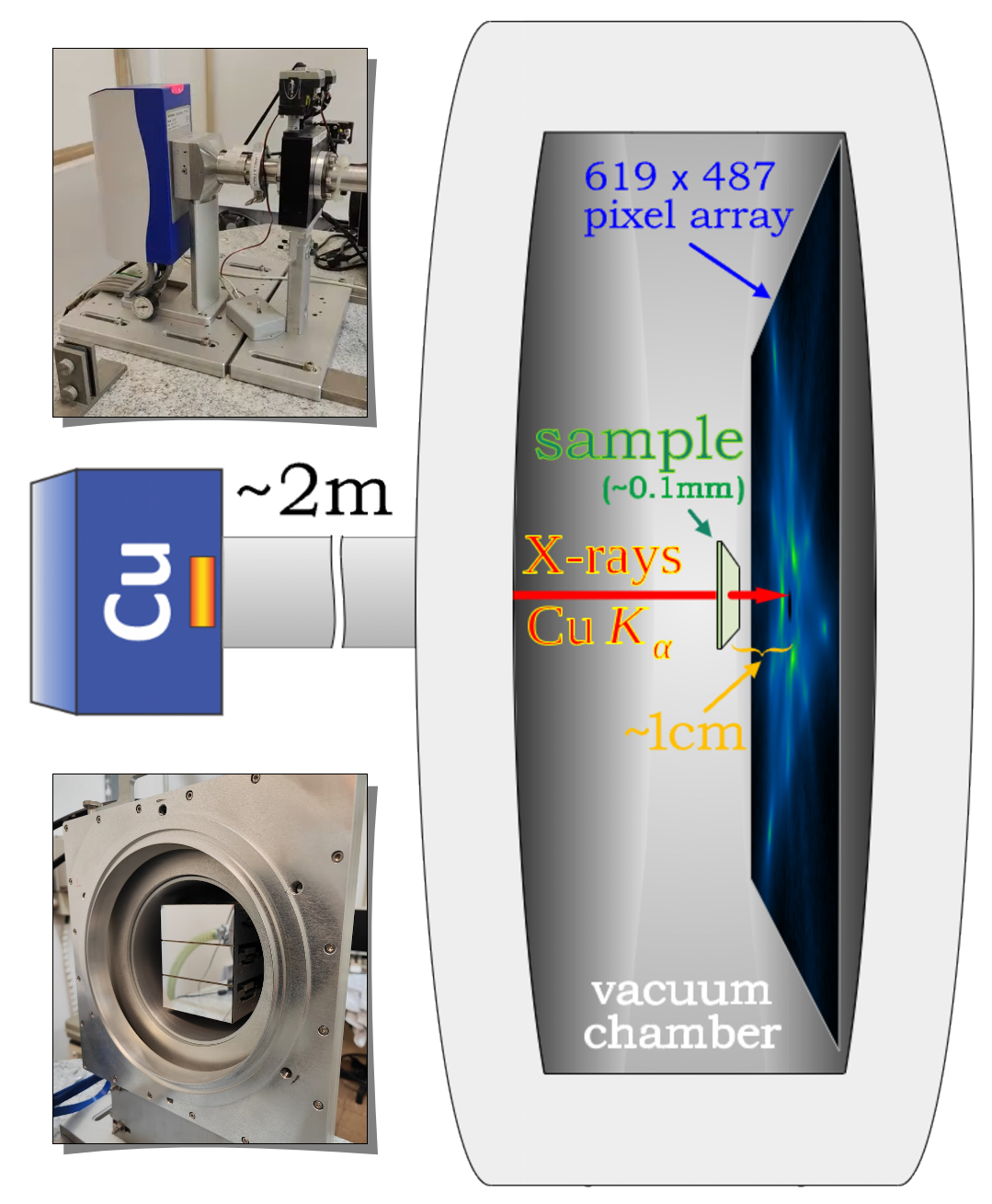}
    \caption{Integrated vaccum geometry of the experimental setup: X-ray microsource and optics (top inset), beam path, sample stage, and detector (bottom inset).}
    \label{fig:instSetup}
\end{figure}

X-ray data were acquired by using a standard SAXS equipment where the microsource (Xenocs) with Cu target and confocal optics are attached to the sample stage and area detector (Pilatus 300K) in vaccum as schematized in Fig.~\ref{fig:instSetup}.   

\section{Conclusions}

In summary, we have demonstrated that quantitative, momentum-resolved lattice dynamics characterization can be achieved using a standard laboratory benchtop X-ray instrument. By coupling transmission TDS under an integrated vacuum geometry with global Differential Evolution optimization, we successfully extracted interatomic force constants up to the 5th-nearest-neighbor shell for single-crystal silicon. The resulting phonon dispersion relations show remarkable fidelity to benchmark data obtained from large-scale Inelastic Neutron Scattering facilities across the full Brillouin zone, demonstrating the robustness of the methodology even in the presence of realistic structural imperfections and sample preparation constraints. 

Beyond validating bulk force-constant formalisms in model semiconductors, this benchtop approach circumvents the severe sample-volume demands of INS ($\sim\text{cm}^3$) and the strict facility scheduling of synchrotrons. Providing full-zone phonon sensitivity with thin, sub-millimeter specimens in an accessible laboratory environment establishes a powerful pathway to directly benchmark DFT exchange-correlation functionals, investigate vibrational anisotropy, and probe lattice dynamics in emerging quantum, thermoelectric, and optoelectronic materials.

\begin{acknowledgments}
This work was supported by the São Paulo Research Foundation (FAPESP) under Grant No.~2023/10775-1. P.~H.~V.~S. and R.~F.~S.~P. acknowledge financial support from FAPESP under Grant Nos.~2025/01027-7 and 2021/01004-6, respectively.
\end{acknowledgments}

\bibliography{tds_manuscript}

\appendix
\onecolumngrid 
\section{Supplementary Information}
\label{sec:si}

\renewcommand{\thefigure}{A\arabic{figure}}
\setcounter{figure}{0}

\begin{figure}[h]
    \includegraphics[width= 0.4\textwidth]{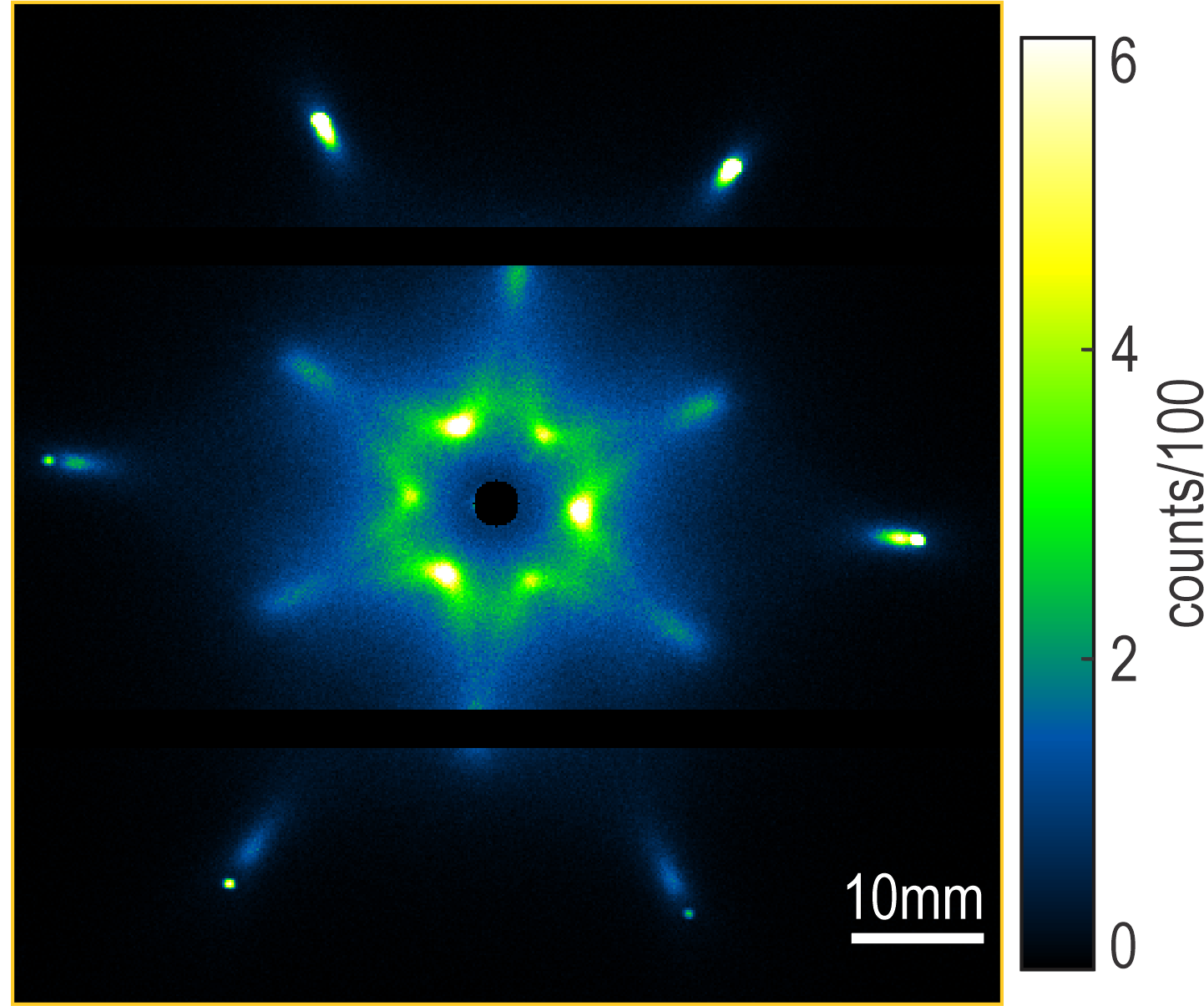}
    \caption{Experimental TDS pattern of the Si(111) sample, a dendritic-web silicon for solar cell applications \cite{Morelhao2000}, collected after vertical sample translation to a position where the incident beam probes a twinned region.}
    \label{fig:sixfold}
\end{figure}

\begin{figure}
    \includegraphics[width= 0.95\textwidth]{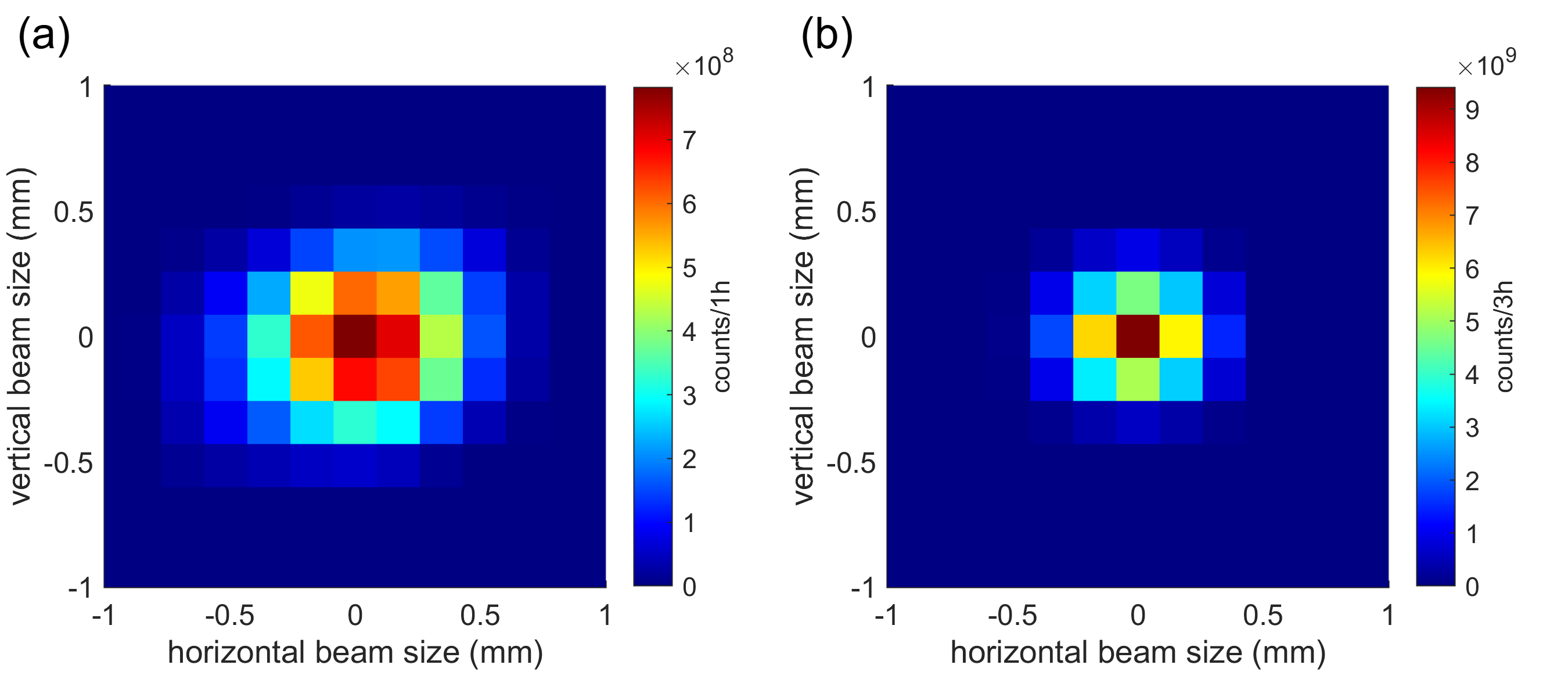}
    \caption{Direct beam profiles recorded at the detector position during acquisition of the TDS images in Figs.~\ref{fig:tdsnew}a and \ref{fig:tdsnew}f.}
    \label{fig:directbeam}
\end{figure}

\begin{figure}
    \centering
    \includegraphics[width=0.4\textwidth]{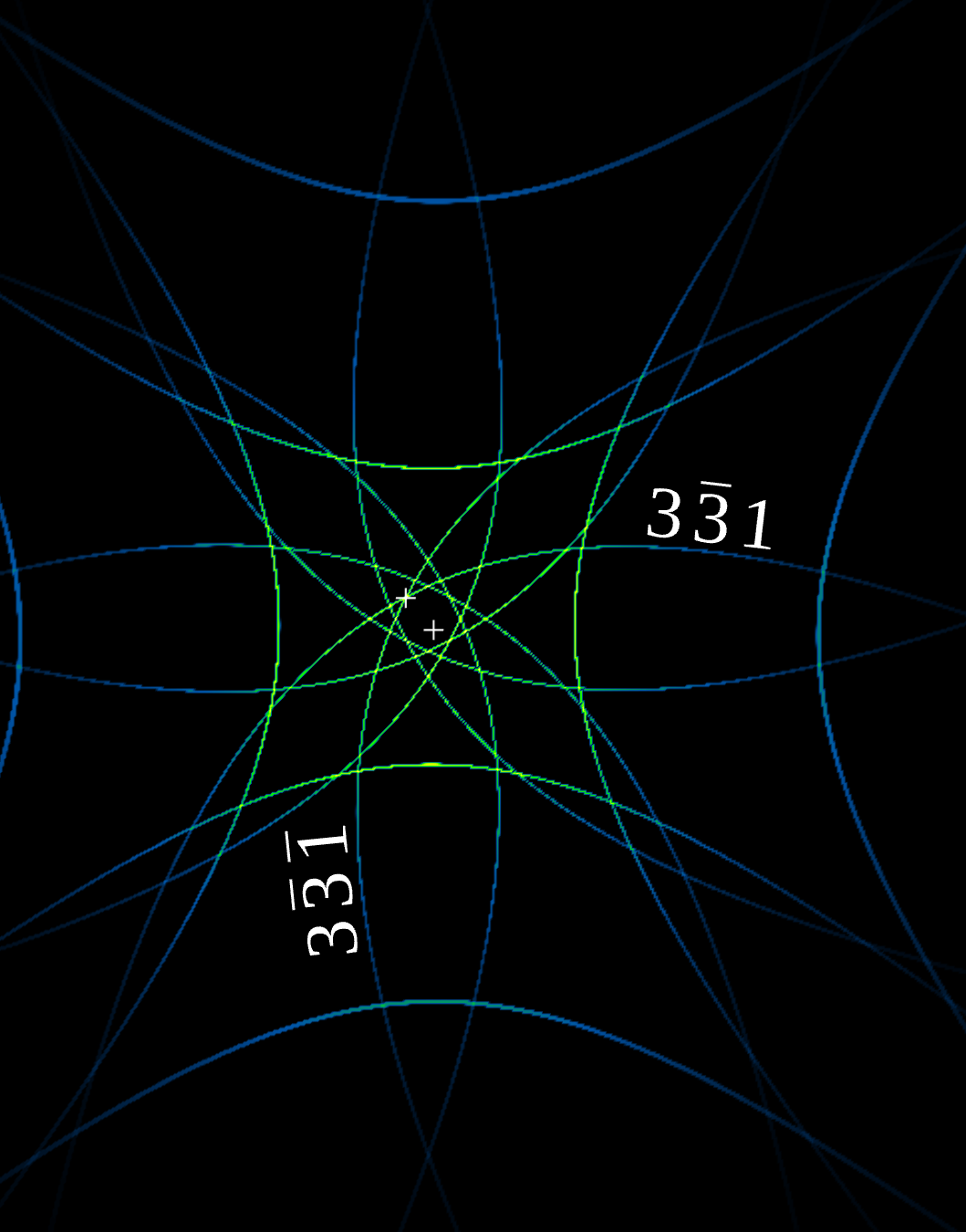}
    \caption{Analysis of Diffuse Multiple Scattering (DMS) features in Si (100). Positions of the $\{331\}$ family of DMS lines on the entire detector area, mapped for the specific acquisition geometry of Fig.~\ref{fig:tdsnew}f through the framework of Estradiote \textit{et al.} (\citeyear{Estradiote2025}).}
    \label{fig:dmslines}
\end{figure}

This Supplementary Information section provides complementary experimental characterizations and geometric simulations related to the findings discussed in the main text. Specifically, we present: (\textit{i}) the experimental TDS pattern collected from a twinned domain of the dendritic-web silicon (111) sample (Fig.~\ref{fig:sixfold}), illustrating the transition from pristine three-fold to apparent six-fold rotational symmetry induced by twin-domain overlap; (\textit{ii}) direct X-ray beam profiles measured in both high- and ultra-high-resolution instrumental configurations (Fig.~\ref{fig:directbeam}), defining the spatial beam profile and resolution limits of the benchtop setup; and (\textit{iii}) calculated diffuse multiple scattering (DMS) line trajectories in transmission geometry for the Si (100) crystal orientation (Fig.~\ref{fig:dmslines}), showing the geometric projection onto the 2D detector of lines arising from second-order scattering processes \cite{Estradiote2025}.

\end{document}